\documentstyle[12pt]{article}
\newcommand{\cc}{\cite}

\newcommand{\beq}{\begin{equation}}
\newcommand{\eeq}{\end{equation}}
\renewcommand{\abstract}{}

\def\lagr{\hbox{$\cal L$}}

\def\dalam{\hbox
{\vrule\vbox{\hrule\hbox to 1ex{ \hfill}\kern 1 ex\hrule}\vrule}}

\def\1/2{\hbox{$ {1 \over 2}$ }}
\def\vf{\varphi}
 
\def\p{\psi}

 \def\E{\hbox{$\cal E $}}
\def\ve{\varepsilon}
\def\w{\omega}
\def\mf{m_F}
\def\mps{{\mf^2 R \over \pi}}

\def\pd{\partial}

\def\L{\Lambda}
\def\pb{\bar \psi}

\def\W{\Omega}
\def\z{\zeta}
\def\mf{m_F}
\def\ex{\hbox{e}}

\def\<{\langle}
\def\>{\rangle}

\def\a{\alpha}

\def\g{\gamma}  
  
\def\l{\lambda}   \def\L{\Lambda}
\def\s{\sigma}

\def\c{\chi}

\def\m{\mu}
\def\n{\nu}
\def\t{\tau}

\def\z{\zeta}
\def\w{\omega}
\def\tt{\theta}

\def\tl{\tilde \tau}
\def\tk{\tilde k}
\def\der{{d \over d\tau}}

\def\({\left(}
\def\[{\left[}
\def\){\right)}
\def\]{\right]}

\def\inf{\infty}
\def\pd{\partial}

\title{Calculation of Ground State Energy for Confined Fermion
Fields}

\author{ Igor O. Cherednikov\thanks{Email: igorch@thsun1.jinr.ru;
igorch@goa.bog.msu.ru }
 \\ \\ {\em Laboratory of Theoretical Physics}
\\ {\em Joint Institute for Nuclear Research} \\
{\em 141980 Dubna, Moscow region, Russia} \\ \\ { and} \\ \\
{\em Institute for Theoretical Problems of Microphysics} \\ {\em
Moscow State University} \\ {\em 119899 Moscow, Russia} }

\date{ \ \ \  }

\begin{document}

\maketitle

\begin{abstract}

A method for renormalization of the Casimir energy of confined fermion
fields in $(1+1)D$ is proposed. It is based on the extraction of
singularities which appear as poles at the point of physical value of the
regularization parameter, and subsequent compensation of them by means
of redefinition of the ``bare'' constants.
A finite ground state energy of the two-phase hybrid fermion bag
model with chiral boson-fermion interaction is calculated as the function
of the bag's size.

\end{abstract}

\section{Introduction}
Nowadays a number of methods is developed for
evaluation of the Casimir energy for systems with quantized fields
under nontrivial boundary conditions \cc{rev}. The most known of
them are the Green function method, the zeta function approach,
the application of contour integration, the
multiple scattering method, and the direct mode summation
with thermal regularization \cc{methods} (for recent discussion
see \cc{new}). An important field of
application for these techniques is an investigation of ground
state energy for the quark bag models with account of one-loop
corrections due to the filled Dirac's negative energy sea \cc{bags}.
In this paper we propose an approach for explicit calculation
of ground state energy for the models with confined fermion field coupled
to the boson (scalar) field in a certain spatial region, so called hybrid
bag models \cc{found}. This method allows to extract the singular terms from
the divergent sums of eigenvalues $\w_n$ and obtain the finite Casimir energy
of fermionic sea as the result of absorbing singularities into the
``dressed'' model parameters.
The realization of a such strategy requires, as a rule, an including
of contact terms into the ``classical'' expression for the energy of system.
The form of these terms is determined by their dependence on the geometric
parameters \cc{contact}.
By means of this method, we renormalize the ground state
energy of the hybrid chiral bag model in $(1+1)D$ and study it as a function
of the bag's size. It's shown that this function has unique minimum what means
that there exists a stable configuration for the considered model.

\section{Regularization}

The ground state energy of fermion field is defined as the vacuum
expectation value of the Hamiltonian:
\beq \< 0| H |0 \> \equiv E_0 = - \1/2 \Big( \sum _{\w_n > 0 } \w_n -
\sum _{\w_k < 0 } \w_k \Big) \ .  \eeq Provided that the spectrum $\w_n$
possesses the symmetry $\w_n \to - \w_n$, the vacuum energy is
\beq E_0 = - \sum_{\w_n > 0} \w_n.  \eeq
This sum diverges. On of the ways to regularize it is to use
the exponential cutoff \cc{johnson, pred} what yields
\beq E_0^{exp} (\t) = \lim_{\t \to 0}
\Big(- \sum_{\w _n > 0} \w_n \ex^{-\t \w _n} \Big) = \lim_{\t
\to 0} \der \sum_{\w_n > 0 } \ex^{-\t \w _n},  \eeq
where $\t = \tl \mu^{-1}$ and $\mu$ has the dimension of mass.
This regularization can be applied for calculation of the Casimir energy
in several simple situations
\cc{rev, johnson, pred},
but usually only for systems with an explicitly known spectrum $\w_n$.
In this paper we'll try to show that one can succeed in calculations even if
this is not the case.
Suppose that an asymptotical
expansion of $\w_n$ for $n \geq 1$ \cc{found} can be found for a model with
an unknown spectrum. This expansion reads
\beq \w_n = \sum_{i=1}^{-\inf} \W_i n^i = \W_1 n + \W_0 +
{\W_{-1} \over n} + O\({1 \over n^2}\) \ .  \eeq
These are the three leading terms in (4) which determine
the divergencies in (2).  To begin, consider the case when the
expansion (4) contains only two terms (what really takes place, e. g.
for the free massless fermion field in (1+1)-bag):
\beq \w_n^{(1)} = \W_1 n + \W_0 \ .  \eeq Then the regularized energy is
\cc{johnson} \beq E^{exp}_1(\t) = \lim_{\t \to 0} \der \sum_{n=1} \ex^{-\t (\W_1 n + \W_0)} - \w_0 =
\lim _{\t \to 0} \der \Bigg( \ex^{-\t \W_0} \s_1(\t)\Bigg) -  \w_0 \ , \eeq
где \beq \s_1(\t) = \sum_{n=1} \ex^{-\t \W_1 n } = {1 \over \ex^{\t
\W_1} - 1} \ .  \eeq The term $\w_0$ in (6) is written separately because
the expansion (4) is not valid for it.
By virtue of \beq {1 \over
\ex^{x} - 1} = \sum _{k=0} {B_k \over k!} x^{k-1} \ , \eeq where $B_k$
are the Bernoulli numbers, one gets
\beq E^{exp}_1(\t) = - {1 \over \t^2 \W_1} +
{\W_1 \over 12} + {\W_0 \over 2} + {\W_0^2 \over 2\W_1} - \w_0  \ . \eeq
This expression contains the quadratic divergence
$E^{quad} (\t)
= - {1 \over  \t^2 \W_1} = - {\m^2 \over \tl^2 \W_1} $  depending
on the arbitrary mass $\m$ and the geometric parameters of the bag from $\W_1$.

Consider now the case when
\beq \w_n^{(2)} = \W_1 n + \W_0 +
{\W_{-1} \over n} \ .  \eeq In this situation the regularized energy is
\beq E^{exp}_2  (\t) = \lim_{\t \to 0} \der \s_2 (\t) -
\w_0 \ , \eeq где \beq \s_2 ( \t ) = \sum_{n=1} \ex^{-\t \Big(
\W_1 n + \W_0 + {\W_{-1} \over n} \Big) } =\ex^{-\t \W_0} \sum_{n=1}
\ex^{-\t \W_1 n } \ex^{-\t \W_{-1}/n} \ . \eeq
One easily see that only the two leading terms in the expansion
of $\ex^{-\t \W_{-1}/n}$ in powers of $\t$ yield a non-vanishing contribution
to $E^{exp}_2 (\t)$ as ${\t \to 0}$. Then we get
$$ \s_2 (\t) = \ex^{-\t \W_0} \sum_{n=1} \ex^{-\t \W_1 n}
\(1 - {\t \W_{-1} \over n } \) + O(\t ^2) = $$ \beq = \s_1 (\t) + \t
\W_{-1} \ex^{-\t \W_0} \sum_{n=1} {1 \over n} \ex^{-\t \W_1 n} +
O(\t ^2)\ , \eeq where $\s_1 (\t)$ is already obtained in (7).
Using the known relation
$$ \sum_{n=1} {1 \over n}
\ex^{- \a n} = - \ln (1 - \ex^{- \a})  $$ we find \beq \s_2 (\t) =
\s_1 (\t) +  \t \W_{-1} \ex^{-\t \W_0} \(\ln \t \W_1 - {\t \W_1
\over 2} \) + O(\t ^2) \ .  \eeq So the regularized energy reads
\beq E^{exp}_2(\t) = -{1 \over \t^2
\W_1} + \W_{-1} \ln \t \m + \W_{-1} \ln {\W_1 \over \m} + {\W_1 \over
12} + {\W_0  \over 2} + {\W_0^2 \over 2\W_1} + \W_{-1} - \w_0.
\eeq
The contribution of the terms of order $O(1/n^2)$ ($E^{fin}$) is finite
and can be found in any particular case. The divergent parts
$$E^{quad}(\t) =
-{1 \over \t^2 \W_1} \ \ \hbox{and}\ \  \ E^{log}(\t) = \W_{-1} \ln \t \m $$
is to be compensated using a certain renormalization scheme, which will
be discussed later.

Let us compare the result obtained above
with another method of regularization --- the $\z$-function approach.
In this case the energy is regularized as:
\beq E^{zeta}_{1}(s) = - \sum_{n} \w_n \to - \lim_{s \to -1}
\mu^{1+s} \sum_{n} \w_n^{-s} \ , \eeq where the arbitrary mass $\m$
is included in order to restore the correct dimension. We assume that
this mass equals $\m$ in (16).
For $\w_n^{(1)} = \W_1 n + \W_0 $ one has
\beq  E_1^{zeta} (s) = - \lim_{s \to -1} \m^{s+1} \sum_{n=1}
(\W_1 n + \W_0)^{-s} - \w_0 = -\lim_{s \to -1} z_1 (s) - \w_0 \
, \eeq where $$ z_1 (s) = \m^{s+1} \sum_1 (\W_1 n +
\W_0)^{-s} = $$ \beq = \sum_1 {1 \over (\W_1 n)^s}\(1 - {s \W_0
\over \W_1 n} + {s(s+1) \over 2!} \({\W_0 \over \W_1 n}\)^2  +
O(1/n^3) \) \ .  \eeq It is clear that the contribution of terms
$O(1/n^3)$ vanishes as $s \to -1$, so the regularized sum
$E_1^{zeta} (s)$ is determined by the leading three terms
in the expansion (18).
Then one finds:  \beq  z_1 (s) = \m^{s+1} \(\W_1^{-s} \z(s) - {s \W_0 \over \W_1^{s+1}} \z
(s+1) + {\W_0^2 \over 2 \W_1^{s+2}} s(s+1)\z (s+2) \) . \
\eeq The values of $\z (z)$ analytically continued to the total real
axis are known \cc{dr_watson}: \beq \z (0) = - {1 \over 2} \
, \ \ \ \z (-1) = - {B_2 \over 2} = - {1 \over 12} \ .
\eeq For $\z (z)$ in the vicinity of $1$ we have \beq \lim_{z \to 1}
\z (z) = {1 \over z - 1} + C \ , \eeq where $C=
0.5772156649...$ is the Euler constant.
Thus taking the limit $s \to -1$ one obtains
\beq E_1^{zeta} (s) = {\W_1 \over 12} +
{\W_0 \over 2} + {\W_0^2 \over 2\W_1}- \w_0 \ .  \eeq
This expression reproduces explicitly the finite part of (9).
The absence of the divergent term is due to the analytical continuation
for $\z (z)$.
Taking into account the next term of the expansion in powers of
$1/n$ (10) we find
$$ E_2^{zeta} (s) = $$ \beq = - \lim_{s \to -1} \m^{s+1}
\sum_{n=1} \(\W_1 n + \W_0 + {\W_{-1} \over n}\)^{-s} - \w_0 =
-\lim_{s \to -1} z_2 (s) - \w_0 \ , \eeq where \beq z_2 (s)
= z_1 (s) - {s \W_{-1} \m^{s+1} \over \W_1^{s+1}} \z (s+2) \
\eeq (we write down only the terms which yield a non-vanishing contribution).
The second term in (24) can be found using (21) and the expansion
$x^{\ve } = 1 + \ve
\ln x  + O(\ve^2) $, $\ve = s+1$, $\ve \to 0$.  Then $$\lim_{s \to
-1} s \W_{-1} \({\m \over \W_1}\)^{s+1} \z (s+2) = $$ \beq  = \W_{-1}
\lim_{\ve \to 0 } (-1 + \ve) \(1 - \ve \ln {\W_1 \over \m}\) \({1
\over \ve} + C\) = -\W_{-1} \({1 \over \ve} -1 + C - \ln {\W_1 \over
\m}\) \ .  \eeq Therefore the regularized energy is of the form
\beq E_2^{zeta} (s) = {\W_1 \over 12} + {\W_0 \over 2} +
{\W_0^2 \over 2\W_1} - \w_0 + \W_{-1} \(\ln {\W_1 \over \m} + 1\) -
\W_{-1} \({1 \over \ve} + C\) \ , \eeq
what coincides with (15) (except of the quadratic singularity $E^{quad}_2$)
provided that \beq {1 \over \t} = \m \g \ex ^{1/\ve}  \ ,
\ \hbox{ln} \g = C.  \eeq
If one needs to take into account the contribution of terms
of order $O\({1 \over n^2}\)$, then their sum $(- E^{fin})$ should be
added to (15) and (26).
Therefore it is shown that the exponential and $\z$-function regularizations
provide the equivalent results for regularized Casimir energy in $(1+1)D$.

The singular part is:
\beq E^{div}(\t) = - {1 \over \t^2 \W_1}
+ \W_{-1} \ln \t \m \ .  \eeq It can be removed, for example,
by means of the redefinition of the ``bare'' constants from the initial
Lagrangian. It is very important to note that the way of renormalization
is prescribed
by the dependence of the extracted singularities on the geometric
parameters (in our case the only geometric parameter is the bag's size $R$).
We collect all singularities with the similar dependence from $R$, and then
find the ``contact term'' to be redefined \cc{contact}.

Let us describe this procedure in the simplest case, i. e., MIT bag model
with massive fermions in $(1+1)D$ \cc{johnson, mit}.  The Lagrangian reads
\beq \lagr _{MIT} = \tt (|x|< R)
(i\pb \hat \pd \p - m_F \pb \p - B) + \tt (|x| > R)  (i\pb \hat \pd
\p - M_F \pb \p)  \ , \eeq where $B$ is the so-called bag constant
which characterizes an excess of the energy density inside a hadron
compared to the energy of the nonperturbative vacuum.
Taking the limit $M_F \to \inf$ in the exterior region $|x|>R$ one
gets the ``bag'', which is just the segment of real axis  $[-R, \ R]$.
The boundary conditions
\beq (\pm i \g^1 +1)\p (\pm R) = 0 \eeq lead to the spectrum
\beq \w_n = \sqrt{
\( {\pi \over 2R} n + {\pi \over 4R} \)^2 + \mf^2} \ .  \eeq
Assume that the fermion mass $\mf$ is small and expand the energy
in powers of it. Then (with the accuracy up to $\mf^4$) one has
\beq \w_n = (\W_1n + \W_0) + {\mf^2 \over 2(\W_1n +
\W_0)} + O(\mf^4) \ , \eeq  $\W_1 = {\pi \over 2R}$, $\W_0 = {\pi
\over 4R}$.  For $n \geq 1 $ the expansion (4) can be found, where  \beq
\W_{-1} = {\mf^2 \over 2\W_1} = {\mf^2 R \over \pi} \, \eeq and for
$n=0$ we have \beq \w_0 = \W_0 + {\mf^2 \over 2\W_0} = \W_0 + 2\W_{-1} \ .
\eeq The divergent part of the energy ( according to (28))  is
\beq E^{div}(\t | R)
= \( - {1 \over \t^2 \pi } + {\mf^2 \over 2\pi} \ln \t \m \) 2R = 2
B'(\t) R \ . \eeq In accordance with the approach of \cc{contact, johnson},
the renormalization is performed by means of the redefinition of
the bare bag constant $B$ in the Lagrangian (29):
\beq B= B_0  - B'(\t)
\ .  \eeq  The remaining parts of energy (15) are finite and can be
found explicitly.
The contribution of terms $O\(n^{-2}\)$ in the expansion (4)
is determined by $$ E^{fin} (R) = - {\mf^2 \over 2}
\sum_{n=1} \( {1 \over \W_1 n + \W_0} - {1 \over \W_1n} \) = $$
\beq = {\mf^2 \over 2 \W_1} \sum_{n=1} {1 \over n(2n+1)} = {2 \mf^2 R
\over \pi} (1 - \ln 2)\ .  \eeq Therefore, the renormalized energy
of the fermionic sea for $(1+1)D$ massive MIT bag model as the function
of the bag's size $R$ and renormalized bag constant $B_0$ reads (up to the terms
of order $\mf^4$):
\beq E_{MIT}(R) = 2B_0 R
- {\pi \over 48R} + \mps \(1 - 2\ln 2 + \ln {\pi \over 2R\m} \) +
O(\mf^4) \ .  \eeq Taking the limit $\mf \to 0$ (massless MIT bag model),
one gets from (38) the well-known result of \cc{class}.
Note, that this configuration is unstable and tends to
$R \to 0$. It is possible to make it stable by adding one valence
fermion into the lowest level $n=0$. Then the energy reads
\beq \tilde E_{MIT} (R) = 2B_0 R + {11 \over 48}{\pi \over R}  + \mps
\(3 - 2\ln 2 + \ln {\pi \over 2R\m} \) + O(\mf^4) \ .  \eeq

\section{Two-phase Hybrid Bag Model}

Now we are ready to study the hybrid bag model, in which the fermion
(``quark'') field interacts with scalar (``meson'') field.
Compared to the models considered in \cc{found}, here is no
phase of the massless fermions.
The Lagrangian reads
$$\lagr = i\pb \hat \pd \p + \1/2 \pd_{\mu} \vf \pd^{\mu}\vf -
\tt(|x|<R) \( \1/2 \mf [\pb, \ex^{i \g_5 \vf} \p] - B \) -$$ \beq  -
\tt(|x|>R) ( V(\vf) + \1/2 M_F [\pb, \ex^{i \g_5 \vf}\p ] \ ). \eeq
The commutator in the fermionic forms provides the charge-conjugation
symmetry. For the sake of simplicity, we consider the scalar field
$\vf(x,t)$  in the mean-field approximation (MFA) \cc{chircoupl}, i. e.,
assume it to be a $c-$number function of the space-time variables;
suppose also that it is independent of the temporal coordinate:
$\vf(x,t) = \vf(x)$.
Besides this, we will take into account only the leading terms
of the perturbative expansion
in the chiral coupling constant \cc{chircoupl}. In other words,
the fermionic mass $\mf$ is assumed to be small inside the segment $|x|<
R$ while the mass $M_F$ in the exterior region $|x|>R$ is infinitely large.
Thus the fermion field at $|x|>R$ vanishes and the equations of
motion for $|x|<R$ are
\beq (i \hat \pd - \mf \ex^{i \g_5 \vf}) \p(x,t) = 0, \eeq
\beq \vf ^{''} = i{\mf \over 2} \<\[\pb, \g_5 \ex^{i \g_5 \vf} \p
\]\>_{sea}, \eeq
where in r. h. s. of (42) the v. e. v. of the axial current is taken
accordingly to MFA.
For $|x|>R$, the scalar field is determined by the non-linear equation
\beq -\vf''(x) = V'_{\vf} (\vf) \ . \eeq The solution of the
equations (41, 42) have been firstly found by Sveshnikov and Silaev and studied
in detail in \cc{found}, so we will discuss it here only briefly.
Let us assume that the scalar field is an odd soliton function and for
$|x| > R$ it has the form:
\beq \vf (x) = \pi \(1 -  A \ex^{-m x } \), \ x > 0 \eeq where
$m$ is the meson mass, and $\vf (x)$ for $x < 0$ is determined by the oddness.
The boundary conditions are $$ (\pm i\g^1 + \ex^{i\g_5
\vf(x)} ) \p (\pm R) = 0 \ , $$ $$ \vf(\pm R \pm 0) = \vf(\pm R \mp
0) \ , $$ \beq \vf'(\pm R \pm 0) = \vf'(\pm R \mp 0) \ .  \eeq
The fermionic spectrum possesses the symmetry
$\n \to -\n$, where $\n^2=k^2+\mf^2 $.
The corresponding unitary transformations of the wave function
are $\c \to i\g_1 \c$ (here $\c = \ex^{i\g_5 \vf /2} \p$),
therefore the v. e. v. of the axial current in the r. h. s. of (42) is equal
to zero and the solution of (42) for
$|x| < R$ appears to be the linear function $\vf(x) = 2 \l x$
(for detailed discussion of the self-consistent solution for this system
see \cc{found}).
The eigenvalues $\w_n$ can be obtained from the equation
$$ \(1 - \ex^{ 2 i k R} {\mf + i (\n + k) \over \mf + i(\n - k)}\)
\(1 - \ex^{- 2 i k R} \( {\n - k \over \n + k} \)  {\mf - i(\n + k) \over
\mf - i(\n - k)} \) = $$ \beq =  \( 1 - \ex^{- 2 i k R} {\mf-i(\n +
k) \over \mf - i(\n - k)} \) \(1 - \ex^{-2 i k R}  \( {\n - k \over \n +
k} \) {\mf + i(\n + k) \over \mf + i(\n - k)} \), \eeq where
$\n = \w -\l$.
Provided that the signatures of $\w_n$ and $\n_n$ are the same for all $n$,
the fermionic Casimir energy is determined by (2) with the replacement
$\w_n \to \n_n$.
For the sake of being definite we regularize this expression
by means of the exponential cutoff (3):
$$ E_0^{exp} (\t) = \lim_{\t \to 0}
\Big(- \sum_{\n _n > 0} \n_n \ex^{-\t \n _n} \Big) = \lim_{\t
\to 0} \der \sum_{\n_n > 0 } \ex^{-\t \n _n} \ . $$
The equation (46) can be written as
\beq \mf \sin 2 R k + k \cos 2 R k  = 0 \ . \eeq
Using the expansion in powers of $\mf$ for $k$
\beq k = \tk_0 + \mf \tk_1 + \mf^2 \tk_2
+ O(\mf^3) \  \eeq one obtains \beq \n = \tk_0 + \mf \tk_1 + \mf^2 \(
\tk_2 + {1 \over 2 \tk_0} \) + O(\mf^3) \ .\eeq
One can solve (47) in any order, what yields (up to $O\(m^3\)$),
\beq \n_n = {\pi \over 2R}n + {\pi \over 4R} + {2 \over \pi} \mf (1+\mf R)
{1 \over (2n+1)} - {16 R \mf^2 \over \pi^3 (2n+1)^3} + O(\mf^3) \ .
\eeq Expanding $\n_n$ in powers of $1/n$ one gets  \beq \n_n = \W_1 n +
\W_0 + {\W_{-1} \over n} + O\({1 \over n^2}\) \ , \eeq where \beq \W_1
= {\pi \over 2R} , \ \W_0 = {\pi \over 4R} , \ \W_{-1} = {\mf \over
\pi} (1+\mf R) \ , \eeq
what differs from the massive MIT bag model by the term
$\mf/\pi$ in $\W_{-1}$ which is independent of the bag's size $R$.
For $n=0$ we have \beq \n_0 = \W_0 + 2 \W_{-1} - {16 R
\mf^2 \over \pi^3 } \ .  \eeq
Now one can use the formulae for the regularized energy (15).
The finite fermionic energy should be of the form
\beq E_F = E_{Cas} + 2 B R + \L \ , \eeq
where $B$  and $\L$ are the bag and cosmological constants,
and $E_{Cas}$ is the divergent fermionic Casimir energy.
Hence the renormalization requires the redefinition of $B$ (36) and $\L$:
\beq \L = \L_0 - {\mf \over \pi} \ln \t \m \ . \eeq The contribution $E^{fin}$
of the non-singular terms $O\({1 \over n^2}\)$ reads
$$E^{fin} = \W_{-1} \sum_{n=1} {1 \over n (2n +1)} + {16 R \mf^2 \over
\pi^3 } \sum_{n=0} {1 \over (2n+1)^3} = $$ \beq = \W_{-1} 2 (1 - \ln
2) + {16 R \mf^2 \over \pi^3 } A \ , \eeq where $A = 1.051799...$.
Therefore the finite energy  of the fermionic sea is
(provided that $\L_0 = 0$):
$$E (R) = 2 B_0 R - {\pi \over 48 R} + $$ \beq + {\mf \over \pi}
(1+\mf R)\[\ln {\pi \over 2 \m R} + 1 - 2 \ln 2 \] +  {16 R \mf^2
\over \pi^3 } A \ .  \eeq

A total energy of this system contains also the contribution of
the scalar field. In order to find it we use the boundary condition
for $\vf$ and it's derivative (45) what yields \beq 2\l = {\pi m
\over m R+1} \ . \eeq By virtue of the virial theorem
in the external region $|x|>R$ one obtain
$\1/2 \vf'^2 (x) = V (\vf)$, so the scalar field energy reads
\beq E_{\vf} (R) = \1/2 \int^{R}_{-R} \! dx \vf'^2(x) +
\( \int^{-R}_{-\infty} + \int^{R}_{\infty} \) \! dx \vf'^2 (x) = {\pi^2 m
\over mR+1} .  \eeq Note that the representation of the
scalar field in the form (44) is valid only at the distances much larger than
the soliton size, i. e. of order $m^{-1}$.
This restriction allows us to postulate (in the framework of our model)
the following relation between the meson mass
$m$ and the bag's radius: $$m R = K_{\pi} \ , K_{\pi}
\geq 1 \ . $$ Thus the contribution of the scalar field to the total energy
of the bag is \beq E_{\vf} (R) = {\pi^2 K_{\pi} \over (1 +
K_{\pi}) R}  \eeq and the total energy consists of the fermionic
and the bosonic parts:
\beq E_{tot} (R) = E_F (R) + E_{\vf} (R) \ . \eeq
In terms of the dimensionless variables the total energy
$\E = E/\mf$ reads $(x = \mf R )$:
$$ \E (x) = 2 B_1 x - {\pi \over 48 x} + $$ \beq + {1 \over \pi} (1 +
x )\[\ln {\pi m_1 \over 2 x} + 1 - 2 \ln 2 \] +  {16 x \over \pi^3 }
A + {\pi^2 K_{\pi} \over (1 + K_{\pi}) x } \ . \eeq
It is easy to see that this energy has the unique minimum and
the configuration is stable.

\section{Conclusion}

The present approach to renormalization of the infinite ground state energy
of the models with confined fermion fields in $(1+1)D$ is based on the
analytical regularization of the divergent sums \cc{analyt}
(e. g., the exponential
cutoff, or $\z$-function regularization), the extraction of the singular
terms in the form of poles at the point of the physical value of regularization parameter,
and subsequent absorbing of these singularities by means of redefinition
of bare constants from the initial Lagrangian.
In this framework
the singularities are isolated unambiguously using the scheme analogous
to MS in QFT.  The remaining parts are finite and demonstrate the non-trivial dependence
on the geometrical parameters of the model. The generalization of the
proposed framework to the higher dimensions appears to be a separate problem
and will be the subject of the future investigations \cc{my_new}.

\section{Acknowledgments}

Author would like to express his gratitude to
Prof. K. A. Sveshnikov for fruitful discussions and critique
and Dr. O. V. Pavlovsky for discussions and many useful remarks.
Author thanks the Abdus Salam ICTP in Trieste for warm hospitality
during 2 month in 1999 when a part of this work was done.
This work is partially supported by RFBR (00-15-96577) and Young scientists Fellowship
of Nuclear Physics Institute, Moscow.

\end{document}